\documentclass[
 reprint,
 amsmath,amssymb,
 aps,
]{revtex4-2}

\usepackage{graphicx}% Include figure files
\usepackage{dcolumn}% Align table columns on decimal point
\usepackage{bm}% bold math
\begin{document}

%\preprint{APS/123-QED}

\title{Controlling chaos: Periodic defect braiding in active nematics confined to a cardioid}% Force line breaks with \\
% \thanks{A footnote to the article title}%

\author{Fereshteh L. Memarian}
\author{Derek Hammar}
\author{Md Mainul Hasan Sabbir}
\author{Mark Elias}
\author{Kevin A. Mitchell}
\email{kmitchell@ucmerced.edu}
 %\altaffiliation[Also at ]{Physics Department, XYZ University.}%Lines break automatically or can be forced with \\
\author{Linda S Hirst }%
\email{lhirst@ucmerced.edu}
\affiliation{%
 Department of Physics, University of California, Merced, 5200 N. Lake Rd, Merced CA 95343 
 % This line break forced with \textbackslash\textbackslash
}%

% \collaboration{MUSO Collaboration}%\noaffiliation

%\author{Charlie Author}
 %\homepage{http://www.Second.institution.edu/~Charlie.Author}
%\affiliation{
% Second institution and/or address\\
% This line break forced% with \\}
%\affiliation{
 %Third institution, the second for Charlie Author}
%\author{Delta Author}
%\affiliation{%
% Authors' institution and/or address\\
 %This line break forced with \textbackslash\textbackslash}

%\collaboration{CLEO Collaboration}%\noaffiliation

\date{\today}% It is always \today, today,
             %  but any date may be explicitly specified

\begin{abstract}

This work examines self-mixing in active nematics, a class of fluids in which mobile topological defects drive chaotic flows in a system comprised of biological filaments and molecular motors. We present experiments that demonstrate how geometrical confinement can influence the braiding dynamics of the defects. Notably, we show that confinement in cardioid-shaped wells leads to realization of the golden braid, a maximally efficient mixing state of exactly three defects with no defect creation or annihilation. We characterize the golden braid state using different measures of topological entropy and the Lyapunov exponent. In particular, topological entropy measured from the stretching rate of material lines agrees well with an analytical computation from braid theory. Increasing the size of the confining cardioid produces a transition from the golden braid, to the fully chaotic active turbulent state.

%\begin{description}
%\item[Usage]
%Secondary publications and information retrieval purposes.
%\item[Structure]
%You may use the \texttt{description} environment to structure your abstract;
%use the optional argument of the \verb+\item+ command to give the category of each item. 
%\end{description}
\end{abstract}

%\keywords{Suggested keywords}%Use showkeys class option if keyword
                              %display desired
\maketitle

%\tableofcontents
Active matter represents a class of materials in which individual sub-units consume locally available energy to create coherent motion at larger scales. They are important materials for elucidating non-equilibrium behavior and for exhibiting features representing, or reminiscent of, biological function.  Examples of active matter exist across length scales, from swarms of animals, e.g. fish~\cite{Katz11}, birds~\cite{Toner95}, or ants~\cite{Buhl06}, to aggregates of eukaryotic cells~\cite{Saw17,Kawaguchi17,Prost15}, bacteria~\cite{Sokolov07,Wensink12,Dunkel13} or synthetic particles~\cite{Palacci13,Yan16,Narayan06}.

In addition to its importance in the biological sciences, active matter has emerged in recent years as a new paradigm for materials design, with the creation of novel phases with nematic order using biological molecules~\cite{Schaller10, Sanchez12, Henkin14,Giomi15, DeCamp15, Guillamat16, Doostmohammadi17}.
Despite a high degree of interest from the physical science perspective, the understanding of fluid dynamics in these active phases is still very limited. This letter focuses on a well-known active fluid, first pioneered by the Dogic group~\cite{Sanchez12}. The active material is formed from an aqueous mixture of microtubules and clusters of kinesin molecular motors, then confined to a two-dimensional (2D) layer. Short, rod-like microtubules at high density in the 2D layer give rise to nematic ordering.  Powered by ATP hydrolysis, neighboring bundles of microtubules slide parallel to each other through the action of the kinesin motors. At certain points in the active fluid, the nematic order may break down, produciing voids as centers for topological defects. These defects have the same symmetry as those observed in traditional liquid crystals, with topological charges of plus and minus one-half.  They are created and annihilated in pairs of opposite charge, so that total charge is conserved.  The defects move around one another in a typically complicated pattern termed active turbulence. In a recent paper~\cite{Tan19}, we introduced the framework that the positive defects act as virtual stirring rods, generating a self-mixing fluid with chaotic advection. The topological braiding of these rods about one another in space-time determines the degree of stretching in the fluid.

\begin{figure}[b]
\includegraphics{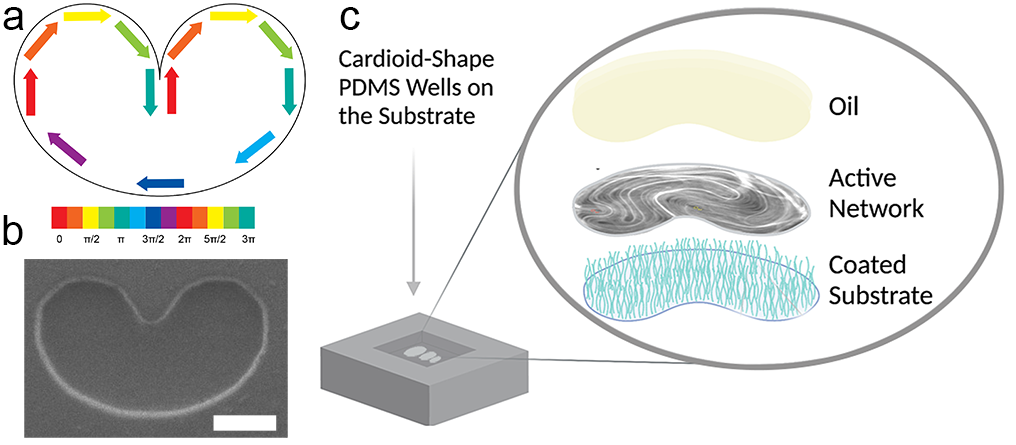}
\caption{\label{fig1} a) The cardioid has a boundary topological charge of 3/2, as illustrated by the arrows.  b) Scanning electron microscope (SEM) image of a PDMS cardioid well before filling. c) Cartoon illustrating the construction of the cardioid wells. }
\end{figure}
 
\begin{figure*}
\includegraphics{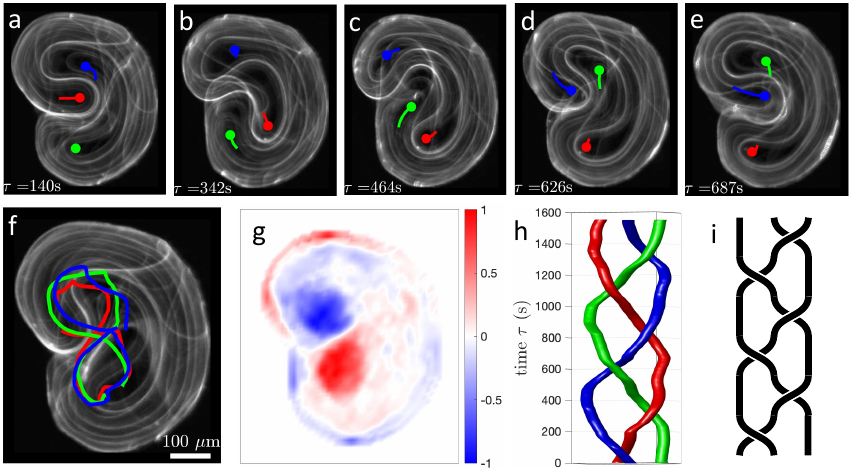}
\caption{\label{fig2} \textbf{Active fluid confined to a cardioid} (a-e) Fluorescence microscope images showing the dynamics of the active fluid confined to a cardioid. Five different time-points from a movie are shown (see supplemental movie S1) in which the microtubules are fluorescently labelled. Three defects are present in all frames and their motion is tracked as shown by the red, green and blue markers. f) Defect trajectories from the same movie are overlaid onto a single frame.  Scale bar is 100$\mu$m. g) Time-averaged vorticity map for the complete movie showing the double gyre flow structure, with positive (red) and negative (blue) vorticities mapped.  Max vorticity set to 1. h) braid diagram for the defect trajectories and i) Cartoon braid diagram for the golden braid.}
\end{figure*}

Controlling the dynamics of active materials is a central goal for the field, and various mechanisms have been proposed for steering defects, such as spatially varying the activity~\cite{Zhang21b}, the oil substrate thickness~\cite{Thijssen21,Khaladj22} and oil type~\cite{Sagues_smectic}. Since the material can be considered a self-mixing fluid, stirred by autonomously moving defects, controlling their braiding (stirring) pattern is a key goal. In this letter, we demonstrate how geometrical confinement can influence the braiding dynamics of the defects.  The key result is that confinement in cardioid-shaped wells can produce the ``golden braid'',  which is the braid of three strands commonly used for braiding hair.  It is a maximally efficient mixing state of three defects, with no creation and annihilation events, that maximizes \emph{topological entropy} per time step~\cite{Finn11}.  Topological entropy, a key measure of chaos adopted from the nonlinear dynamics literature, is simply the asymptotic exponential growth rate in the length of a material curve, when passively advected in a 2D flow.  Ref.~\cite{Tan19} first applied topological entropy to the study of active nematics and showed that topological entropy resulted from ``stirring'' the system only by the positive defects and that the entropy could be computed from the braid type of the defect motion.

When confining an active nematic, the boundary geometry is critical because it determines the relative number of positive $(n_+)$ and negative $(n_-)$ defects in the system.  The boundary geometry is characterized by its topological index, defined as the number of clockwise revolutions made by the tangent vector as the boundary is traversed in the clockwise direction.  For example, a circle, indeed any smooth closed curve, has a topological index of $+1$.  Other topological indices are possible if there are cusps in the boundary, where there is a discontinuity in the curvature (although the tangent vector remains continuous.)  Each cusp pointing toward the interior (or exterior) adds $+1/2$ (or $-1/2$) to the topological index of the boundary.  The cardioid (heart-shape) in Fig.~\ref{fig1}a, for example, has topological index $+3/2$.  The well-known Poincar\'{e}-Hopf Index Theorem implies that the sum of the internal topological charges equals the topological index of the boundary (assuming tangential alignment at the boundary).  Thus, inside a circular boundary, there must always be two more $+1/2$ charges than $-1/2$ charges, with the minimal number of $+1/2$ charges being two.

Experimental results using circular confinement of microtubule-based active nematics have been reported by other groups \cite{opathalage,Hardouin22}.  However, the minimum number of two defects was never observed.  Ref.~\cite{opathalage} came close, with two +1/2 defects winding around one another for long stretches of time.  But this two-defect steady state was unstable, and a third defect would always be created at the boundary and enter the system.  This can be understood as a consequence of the need to produce topological entropy.  Any braiding pattern of two stirring rods generates zero topological entropy, which is inconsistent with the extensile nature of the microtubule system.  Thus, a third defect must ultimately be produced to create topological entropy.  This motivated us to work with the cardioid, whose single interior cusp produces at least three $+1/2$ defects. Our passive control approach, based on engineering the shape of the boundary, thus suffices to ensure the minimum number of defects necessary to create topological entropy. This enables the possibility of a planar state with no defect creation and annihilation.   A similar topological argument applies to the well-known state of four defects on a sphere.  A sphere has topological index +2, and hence four defects is the minimum required by topology and is also sufficient to generate topological entropy~\cite{Smith22b}.  Thus, this state need not exhibit creation and annihilation events, as realized in Ref.~\cite{keber14}. 

Periodic defect motion was first predicted in simulation \cite{Shendruk17} in a 1D channel with periodic boundary conditions. This motion was approximately observed experimentally with some defect creation and annihilation events~\cite{Hardouin19}.

 We designed and constructed cardioid wells of varying sizes to confine the active fluid into a 2D plane with controlled boundary conditions (Fig.~\ref{fig1}b,c). (See Supplemental Material for detailed cardioid shape.)  We first designed and printed a master mold \cite{Memarian2023}, then used the mold to produce cured PDMS substrates. The PDMS substrates are plasma cleaned then coated with an acrylamide polymer brush to promote hydrophilicity and prevent non-specific protein adhesion. Active MIX (an aqueous solution containing motor proteins \cite{Memarian2023}) is pipetted into the cardioid-shaped well, and a layer of oil, approximately 2mm thick, added on top. For these experiments Guanosine-5’-[($\alpha, \beta$)-methyleno]triphosphate (GMPCPP) microtubules are used. GMPCPP is an analog of GTP(Guanosine-5'-triphosphate), and produces short, more rigid microtubules suitable for forming the nematic state. Fig.~\ref{fig1}c shows a schematic of the experimental design in which a nematic layer of condensed microtubules is formed. For detailed methods see Memarian et al~\cite{Memarian2023}. To visualize microtubule flows, 10~mol\% fluorescent Alexa 488 labelled tubulin is incorporated in the polymerization step. All imaging was performed with a Leica epifluorescence microscope and an ORCA - Flash4.0 LT+ Digital CMOS camera (Hamamatsu Inc.).

Figure~\ref{fig2}a-e shows five snapshots of the active nematic fluid confined in a cardioid. These images, and supplementary video 1, demonstrate the persistence of three +1/2 defects (colored dots), with no additional creation or annihilation events.  The tails on each dot show the defects' recent trail.  Between frames a and c, the green and red defects have swapped positions, and between frames c and e, the blue and green defects have swapped positions. We noted that for small wells, the flow velocity decreased exponentially in time, due to the depletion of ATP.  To correct for this, we exponentially rescaled physical time $t$ into an effective time $\tau$.  That is, we measured $\tau$ with a clock that slowed down at the same rate as the flow (see supplemental material for details.)    Figure~\ref{fig2}f shows the three full trajectories, forming a figure-eight pattern.  These trajectories are created by a double gyre (Fig.~\ref{fig2}g), with defects jumping back and forth between the vortices.  Figure~\ref{fig2}h shows the space-time braiding of the defects about one another.  Importantly, \emph{this braid is topologically identical to the  golden braid} (Fig.~\ref{fig2}i).  Thus, when sufficiently confined to a cardioid, the system naturally realizes the periodic state of maximal topological entropy per swap, among all possible dynamics of three defects.

\begin{figure}
    \includegraphics{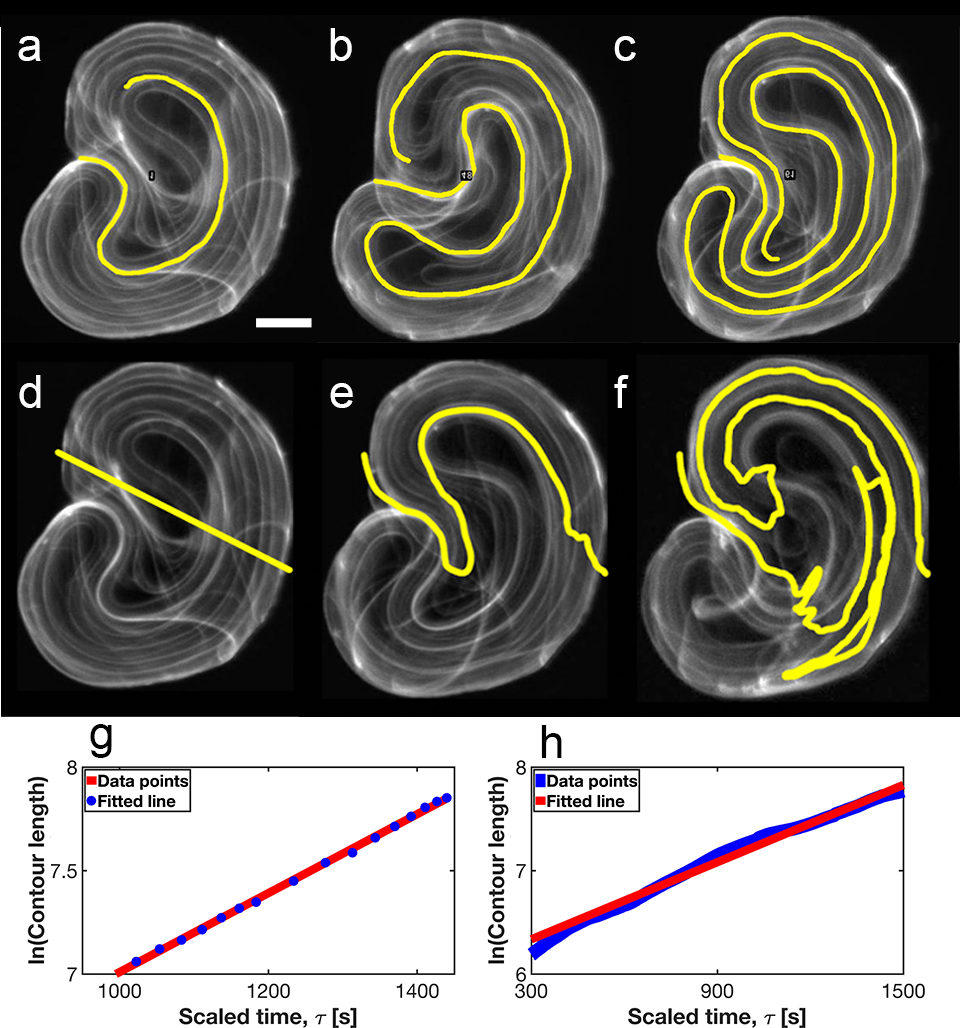}
    \caption{\label{fig3} Representative fluorescence images of active fluid confined to a cardioid at different time points (see supplemental movie 1), overlaid with (a-c) a manually tracked stretching curve and (d-f) a computationally stretched and advected band. g) Semilog plot showing the length of the manually tracked fluid stretching curve in a-c as a function of time, h) Semilog plot of the length of the stretched and advected band in d-f as a function of time.}
\end{figure}

The golden braid dynamics suggests a nontrivial quantitative relationship.  For a system stirred in the golden braid pattern, the minimum topological entropy per swap is provably $h_{\text{braid}} = \log(\phi) = 0.481$, where $\phi = (1 + \sqrt{5})/2$ is the golden ratio~\cite{Finn11}.  Since there are no other obvious sources of mixing beyond the three ``rods'' (i.e., defects), this minimum might in fact be the full topological entropy of the flow.  Furthermore, stretching appears relatively homogeneous throughout the fluid; there are, for example, no coherent Lagrangian vortices that capture fluid for long periods of time and inhibit mixing.  We thus suspect that the stretching rate of a typical material line over a finite time scale (on the order of the defect swap time) will average to $\log(\phi)$.  To test this hypothesis, we measured the exponential growth of the material line formed by the microtubules lining the boundary of the cardioid together with the material extruded into the cardioid interior at the cusp (the bright curve folding back and forth in Fig.~\ref{fig2}a-e.  We manually measured (in ImageJ) the length of the nematic contour from the cusp to a point of maximal curvature near one of the three defects (Fig.~\ref{fig3}a-c).  We assumed that this point was passively advected by the material.  This length was then augmented by half the perimeter of the cardioid.  (The factor of one-half is because material is extruded from both sides of the cusp.)  Figure~\ref{fig3}g plots the log of length versus time, with slope $1.9 \times 10^{-3} \; \text{s}^{-1}$.  Averaging over two such measurements, we obtained the direct curve stretching estimate of topological entropy $\tilde{h}_{\text{dc}} = (1.75 \pm 0.15) \times 10^{-3} \; \text{s}^{-1}$.  The time per swap is measured to be $T_{\text{swap}} = 260 \pm 18 \; \text{s}$, which yields a dimensionless $h_{\text{dc}}  = \tilde{h}_{\text{dc}} T_{\text{swap}} = 0.455 \pm 0.039$.  The key quantitative result of this paper is that $h_{\text{dc}}$ is equal (within error) to the analytical braid result $h_{\text{braid}}= 0.481$.

We compute two more measures of chaos, both derived from the PIV velocity field.  The first of these uses the velocity field to passively advect an arbitrary initial line over a given time interval.  Figure~\ref{fig3}d-f shows the corresponding snapshots.  The measured (dimensionless) stretching rate (Fig.~\ref{fig3}h) is $h_{\text{PIV}} = 0.286 \pm 0.024$, which is notably smaller than $h_{\text{braid}}$.  This is because the PIV velocity does not fully capture the physical velocity, but tends to underestimate it, an effect well known for active nematics~\cite{Serra23}.  The numerically advected curve is thus not as well stretched as the physical curve.  By the third frame  (Fig.~\ref{fig3}f), the advected curve (yellow) follows the nematic contours fairly well in the upper half of the cardioid, but not in the lower half.  (Compare to  Fig.~\ref{fig3}c.)  

\begin{figure}
\includegraphics{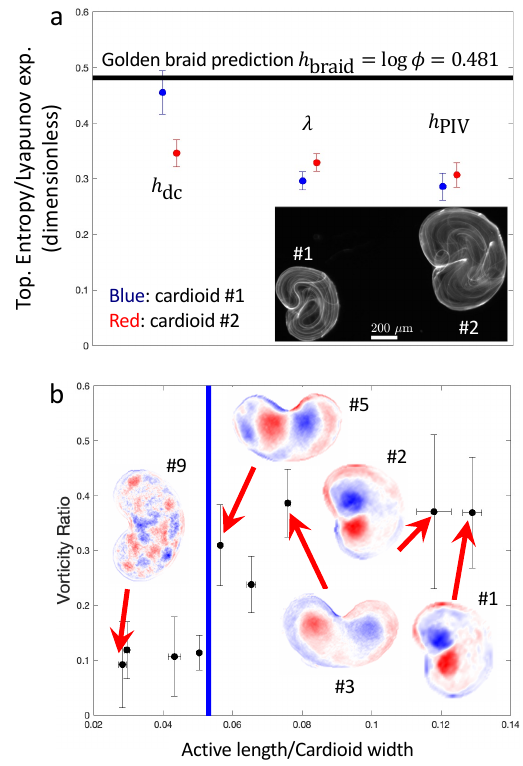}
\caption{\label{fig4} a) Three nondimensionalized measures of chaos for two cardioids (labelled 1 and 2) compared to the braid-theoretic prediction (horizontal line). inset: snapshot of the cardioids imaged on the same substrate (see Supplemental Movie S2).  Scale bar is 200$\mu$m.  b) Ratio of $\langle \langle \omega  \rangle_t \rangle_{\text{rms }r}$ to $\langle \langle \omega \rangle_{\text{rms }t} \rangle_{\text{rms }r}$ plotted for several cardioids of increasing confinement ratio.  The vertical line separates those cardioids with a double-gyre structure (right) from those without (left).  Inset: we show time-averaged images of vorticity for several data points, see also supplemental movies S3 and S4.}
\end{figure}

The final measure of chaos is the Lyapunov exponent, computed as $\lambda = -\langle \mathbf{n}^\perp \cdot (\nabla \mathbf{u}) \mathbf{n}^\perp \rangle$, where $\mathbf{n}^\perp$ is a unit vector perpendicular to the unit director field $\mathbf{n}$ (extracted from the images), $\mathbf{u}$ is the (PIV) velocity field, and the brackets denote the average over space and time.  (See Supplement to Ref.~\cite{Tan19} for the derivation and underlying assumptions.)  We obtain $\lambda = 0.296 \pm 0.017$ in dimensionless units.   (As a check, we compute $\lambda = \langle \mathbf{n} \cdot (\nabla \mathbf{u}) \mathbf{n} \rangle = 0.304 \pm 0.017$, which should be---and is---the same due to net incompressibility.)  The three measures of chaos are graphically represented in Fig.~\ref{fig4}a in blue and compared to $h_{\text{braid}}$ (black horizontal line).  Note that both PIV-derived quantities are equal to one another (within error).  A mathematical theorem in dynamical systems theory states that the topological entropy is always greater than or equal to the Lyapunov exponent (for 2D flows).  The fact that both are equal here is a consequence of the homogeneous nature of mixing throughout the fluid.

We recorded a second, larger cardioid in the same frame as the original.  (See inset of Fig.~\ref{fig4}a and Supplementary movie 2.)  Its flow is faster and closely mimics the golden braid of the smaller cardioid.   However, it is not as faithful, with a few defect pairs being created and destroyed.  This highlights the need to sufficiently confine the material to achieve the exact golden braid pattern.  We  repeat the stretching and Lyapunov exponent analyses, and, when scaled by the swap time $T_{\text{swap}} = 102 \pm 4.9$ s, the PIV-derived results agree with the smaller cardioid (red data in Fig.~\ref{fig4}a.)  However, $h_{\text{dc}} = 0.346 \pm 0.024$ is lower than the golden braid result.  We suspect that this is due to imperfections in the state, in particular the occasional creation of defects at the boundary which should be taken into account to compute the full stretching rate of the boundary curve.  

We obtained data from a series of progressively larger cardioids.  The number of defects quickly increased and defect creation and annihilation events proliferated.  However, the double gyre structure was quite persistent, even when many defects were present.  Figure~\ref{fig4}b shows insets of the integrated vorticity for several examples.  (See supplemental videos for the labelled inset images 1,2,3 and 9.) The images are arranged from left to right with increasing confinement, as measured by the ratio of active length to cardioid width.  The original example (Figs 2,3) is furthest to the right, but all four of the rightmost images show a clear double gyre structure with prominent positive (blue) and negative (red) vorticity domains in the center.  However, the leftmost, least confined cardioid shows no coherent pattern to the vorticity.  Figure~\ref{fig4}b also plots the ratio of $\langle \langle \omega  \rangle_t \rangle_{\text{rms }r}$ to $\langle \langle \omega \rangle_{\text{rms }t} \rangle_{\text{rms }r}$, where the former is the rms spatial average of the direct time-averaged vorticity and the latter is the rms average in both space and time.   The direct time average of vorticity measures the persistence of vorticies in the same location.  It goes to zero when vorticites meander about.  A clear jump in the vorticity ratio occurs at the blue vertical line, demarcating the transition to the persistent double gyre flow.  These results suggest three domains of behavior: strong confinement, when the dynamics closely follows the golden braid; medium confinement, when the golden braid breaks down, but the double gyre is still present; and weak confinement, when the double gyre breaks down and the fully active turbulent state emerges.

In summary, we have demonstrated that by engineering a confining boundary to include a cusp, it is possible to realize the spontaneous self-driven golden braid state in an active nematic flow. The golden braid is a maximally efficient mixing state of exactly three defects with no creation or annihilation events. Exploring different measures of chaotic advection, we demonstrate that the stretching rate of material lines in the active fluid agrees well with the expected analytical computation from braid theory. We further demonstrate via the vorticity that increasing the size of the confining cardioids produces a surprisingly sharp transition from the double gyre---necessary for the golden braid state---to the fully chaotic active turbulent state. These results are significant because they demonstrate the robust connection between fluid mixing theory and active nematics, reinforcing the concept that defects act as virtual stirring rods and that those defects can be represented as a braid. Demonstration of the golden braid, a unique periodic mixing state, opens up new possibilities for more exquisite passive control of motile defects using boundary topology and microfluidics. 

\begin{acknowledgments}
The authors acknowledge support from the National Science Foundation (NSF) awards DMR-1808926, Center of Research Excellence in Science and Technology: Center for Cellular and Biomolecular Machines at the University of California Merced (HRD-1547848) and the Brandeis Biomaterials Facility Materials Research Science and Engineering Center (DMR-2011486). We also thank Dr. Bin Liu at the University of California, Merced for assistance with 3D printing.

\end{acknowledgments}

%apsrev4-2.bst 2019-01-14 (MD) hand-edited version of apsrev4-1.bst
%Control: key (0)
%Control: author (72) initials jnrlst
%Control: editor formatted (1) identically to author
%Control: production of article title (-1) disabled
%Control: page (0) single
%Control: year (1) truncated
%Control: production of eprint (0) enabled
%

%\bibliographystyle{apsrev4-2}
%\bibliography{MyBibDeskBib}

\end{document}